\documentclass[11pt]{article}
\usepackage{amssymb}

\usepackage{epsfig,latexsym}

\def\theequation{\arabic{section}.\arabic{equation}}

\renewcommand{\theequation}{\thesection.\arabic{equation}}

\global\arraycolsep=1pt
\input epsf
\input{tcilatex}
\begin{document}

\font\cmss=cmss10 \font\cmsss=cmss10 at 7pt \hfill \hfill IFUP-TH/03-40


\vspace{10pt}

\begin{center}
{\Large \textbf{\vspace{10pt}CONSISTENT\ IRRELEVANT\ DEFORMATIONS\ OF
INTERACTING\ CONFORMAL\ FIELD\ THEORIES}}

\bigskip \bigskip

\textsl{Damiano Anselmi}

\textit{Dipartimento di Fisica ``E. Fermi'', Universit\`{a} di Pisa, and INFN%
}
\end{center}

\vskip 2truecm

\begin{center}
\textbf{Abstract}
\end{center}

{\small I show that under certain conditions it is possible to define
consistent irrelevant deformations of interacting conformal field theories.
The deformations are finite or have a unique running scale
(``quasi-finite''). They are made of an infinite number of lagrangian terms
and a finite number of independent parameters that renormalize coherently.
The coefficients of the irrelevant terms are determined imposing that the
beta functions of the dimensionless combinations of couplings vanish
(``quasi-finiteness equations''). The expansion in powers of the energy is
meaningful for energies much smaller than an effective Planck mass. Multiple
deformations can be considered also. I study the general conditions to have
non-trivial solutions. As an example, I construct the Pauli deformation of
the IR fixed point of massless non-Abelian Yang-Mills theory with }$N_{c}$
{\small colors and }$N_{f}\lesssim 11N_{c}/2$ {\small flavors and compute
the couplings of the term }$F^{3}${\small \ and the four-fermion vertices.
Another interesting application is the construction of finite chiral
irrelevant deformations of N=2 and N=4 superconformal field theories. The
results of this paper suggest that power-counting non-renormalizable
theories might play a role in the description of fundamental physics.}

\vskip 1truecm

\vfill\eject

\section{Introduction}

\setcounter{equation}{0}


Certain power-counting non-renormalizable theories can be quantized
successfully, for example the four-fermion models in three spacetime
dimensions \cite{parisi} in the large N expansion. Using the procedure of
ref. \cite{mid} it is possible to renormalize quantum gravity coupled with
matter in three spacetime dimensions as a finite theory. A theory that is
not power-counting renormalizable does not necessarily violate fundamental
physical principles and cannot be discarded \textit{a priori}. At present,
it is not clear why some theories can be quantized and other cannot, which
theories are meaningful and which are meaningless. In this paper I\ present
results that are expected to shed some light on this problem.

\bigskip

In four-dimensions pure gravity is finite to the first loop order \cite
{thooftveltman}, but finiteness is spoiled by the presence of matter.
Moreover, gravity is not finite to the second loop order \cite{sagnotti}
even in the absence of matter. In three dimensions the situation is
different. In ref. \cite{mid} I have formulated a quantization procedure to
construct finite theories of quantum gravity coupled with matter in three
dimensions, under the assumption that the matter sector satisfies certain
restrictions. Those ideas do not generalize immediately to four-dimensional
quantum gravity, but admit a number of other interesting four-dimensional
extensions. In this paper I explore a class of such applications, namely the
construction of consistent irrelevant deformations of interacting conformal
field theories.

In general, it is not known how to define \textit{one} irrelevant
deformation, because as soon as one irrelevant term is added to the
lagrangian, renormalization turns on infinitely many other terms, multiplied
by independent couplings. This spoils predictivity at the level of
fundamental field theory (but not at the level of effective field theory).
The goal of this paper is precisely to disentangle the irrelevant
deformations from one another. This result makes it possible to study
``one'' irrelevant deformation, or ``two'' irrelevant deformations, etc., or
all of them together, which is the usual situation. A single irrelevant
deformation is made of an infinite series of lagrangian terms that
renormalize coherently, with a unique renormalization constant, associated
with a dimensionful coupling (the scale). In this sense, the so-deformed
theory is still physically predictive as a fundamental field theory,
although it is not power-counting renormalizable.

\bigskip

The irrelevant couplings are responsible of the non-polynomial structure of
the renormalized lagrangian. The beta functions and renormalization
constants of the irrelevant couplings are polynomial in the irrelevant
couplings themselves. So, on the one side non-renormalizable theories are
complicated, on the other side they are extremely simple. For this reason,
it is possible to work with them.

Consider a conformal field theory $\mathcal{C}$ and its irrelevant
deformations. If $\lambda $ is the coupling constant that multiplies the
irrelevant term \textrm{O}$_{\lambda }$, the beta function of $\lambda $ has
the form \cite{mid}
\begin{equation}
\beta _{\lambda }=\lambda \gamma _{\lambda }+\delta _{\lambda }.
\label{betagel}
\end{equation}
Here $\gamma $ is the anomalous dimension of $\mathrm{O}_{\lambda }$ and
depends only on the marginal couplings of $\mathcal{C}$. Instead, $\delta $
does not depend on $\lambda $ and depends polynomially on a finite number of
other irrelevant couplings.

A structure as simple as (\ref{betagel}) suggests that in a number of cases
it is possible to solve the finiteness equations $\beta _{\lambda }=0$. A
set of non-trivial solutions has been studied in \cite{mid}. However, in
various situations, the finiteness equations admit only the trivial solution
$\lambda =0$, which is just the conformal theory $\mathcal{C}$. To construct
non-trivial deformations in these cases, it is possible to define
``quasi-finite'' theories, i.e. theories that have a unique runnning
parameter, the scale.

Taking appropriate combinations of dimensionful couplings, it is always
possible to organize the set of couplings of a theory into a unique
dimensionful parameter, the scale $\kappa $, with conventional
dimensionality $-1$, plus dimensionless couplings. A \textit{quasi-finite}
theory is a theory whose dimensionless couplings have vanishing beta
functions. The scale is free to run. If the scale does not run, the theory
is \textit{finite}. If there is no scale, the theory is \textit{conformal}.
In this paper I construct finite and quasi-finite consistent irrelevant
deformations of interacting conformal field theories.

\bigskip

Known examples of quasi-finite theories are the mass deformations of
conformal field theories. Consider for example N=4 supersymmetric Yang-Mills
theory in four dimensions (for an introduction to supersymmetry in the
language of superfields, see for example \cite{grisaru}) and denote the
gauge coupling with $g$. Supersymmetry can be softly broken to N=0 giving
masses $m$ to the scalar fields $\varphi $, for example. After this
breaking, the beta function $\beta _{g}$ remains zero, because, by
dimensional considerations, it cannot depend on $m$. On the other hand, the
mass operator $\overline{\varphi }\varphi $ is not finite (its anomalous
dimension is non-vanishing at $g\neq 0$; see for example \cite{n=4}). This
implies that the scale $m$ does run. Therefore, the deformed theory is
quasi-finite. The quasi-finite theories constructed in this paper are a
counterpart, in the irrelevant sector, of the relevant deformations of
conformal field theories.

On the other hand, N=4 supersymmetry can be broken to N=1 with a chiral mass
deformation. This deformation is finite, because of a well-known
non-renormalization theorem. In this paper I construct also finite chiral
irrelevant deformations of superconformal field theories (section 6).

\bigskip

Now I\ sketch the construction of quasi-finite irrelevant deformations. The
set of couplings can be conveniently split into an energy scale $1/%
\widetilde{\kappa }$ and dimensionless ratios $g_{i}$. The beta functions $%
\beta _{i}$ of the dimensionless couplings $g_{i}$ cannot depend on $%
\widetilde{\kappa }$. The beta function of $\widetilde{\kappa }$ is equal to
$\widetilde{\kappa }$ times a function of the $g_{i}$s. Then, it is
consistent to solve the \textit{quasi-finiteness equations}
\begin{equation}
\beta _{i}=0,~\qquad \frac{\mathrm{d}\widetilde{\kappa }}{\mathrm{d}\ln \mu }%
=\beta _{\widetilde{\kappa }}.  \label{quasifiniteness}
\end{equation}
The solutions of the quasi-finiteness equations are, in general, non-trivial
and contain a unique arbitrary parameter besides the marginal couplings of $%
\mathcal{C}$, namely the value of $\widetilde{\kappa }$ at some reference
energy $\overline{\mu }$. It is also correct to view $1/\widetilde{\kappa }(%
\overline{\mu })$ as the definition of the unit of mass.

In the paper, after developing the general approach, I consider a concrete
model, the Pauli deformation of the IR fixed point of massless non-Abelian
Yang-Mills theory with $N_{c}$ colors and $N_{f}\lesssim 11N_{c}/2$ flavors.
I\ study the self-renormalization of the Pauli term and the structure of the
Pauli deformation to the order $\mathcal{O}(\kappa ^{2})$ included, which is
made of the irrelevant terms of dimensionality 6 (four-fermion vertices and $%
F^{3}$). I solve the quasi-finiteness equations and show that the solutions
relate in a unique way the couplings of $F^{3}$ and the four-fermion
vertices to the coupling of the Pauli term. One-loop calculations are
sufficient for these studies.

Multiple deformations can be defined also, where various dimensionful
parameters run independently. Multiple deformations are not sums or
superpositions of simple deformations.

\bigskip

The paper is organized as follows. In section 2 I present the general theory
of finite and quasi-finite irrelevant deformations and study conditions to
have non-trivial solutions. In sections 3, 4 and 5 I\ study the Pauli
deformation of the IR fixed point of Yang-Mills theory coupled with matter.
I solve the quasi-finiteness equations to the second order in $\kappa $ and
first order in the loop expansion. In section 6 I construct the finite
chiral irrelevant deformations of N=2 and N=4 superconformal field theories.
Section 7 contains the conclusions. The appendix collects a number of useful
identities and the field equations.

\section{Consistent irrelevant deformations}

\setcounter{equation}{0}

Consider the set of irrelevant deformations of a conformal field theory $%
\mathcal{C}$ of interacting fields $\varphi $. The classical lagrangian in $%
d $ dimensions has the form
\begin{equation}
\mathcal{L}_{cl}[\varphi ]=\mathcal{L}_{\mathcal{C}}[\varphi ,\alpha
]+\sum_{i}\kappa ^{i}\sum_{I=1}^{N_{i}}\lambda _{iI}\mathcal{O}_{iI}(\varphi
).  \label{ola}
\end{equation}
The $\mathcal{O}_{iI}$ are a basis of (gauge-invariant) local lagrangian
terms with canonical dimensionalities $d+i$ in units of mass. The index $i$
denotes the ``level'' of $\mathcal{O}_{i}$ (irrelevant operators have
positive levels, marginal operators have level 0 and relevant operators have
negative levels) and can be integer or half-integer. The $\lambda _{iI}$
denote a complete set of essential couplings, labelled by their level $i$
plus an index $I$ that distinguishes the couplings of the same level. The
essential couplings are the couplings that multiply a basis of lagrangian
terms that cannot be renormalized away or into one another by means of field
redefinitions \cite{wein}.

The constant $\kappa $ is an auxiliary quantity with dimensionality $-1$ in
units of mass. Every $\lambda $ is dimensionless. I\ assume that the theory
does not contain masses and superrenormalizable parameters (positive-level
couplings). Parameters with positive dimensionalites in units of mass form
dimensionless quantities when they are multiplied by suitable powers of the
irrelevant couplings. These dimensionless combinations are responsible for
unnecessary complicacies, both at the theoretical and practical levels,
because the beta functions do not depend polynomially on them.

The redundancy of the constant $\kappa $ is exhibited by the invariance of (%
\ref{ola}) under the scale symmetry
\begin{equation}
\lambda _{iI}\rightarrow \Omega ^{-i}\lambda _{iI},\qquad \kappa \rightarrow
\Omega \kappa .  \label{scala}
\end{equation}

\bigskip

\textbf{Structure of the beta functions.} The beta function of $\lambda
_{iI} $ transforms as $\lambda _{iI}$ under the scale symmetry (\ref{scala})
and therefore its structure is
\begin{equation}
\beta _{iI}=\sum_{\{n_{jJ}^{iI}\}}f_{\{n_{jJ}^{iI}\}}\left( \alpha \right)
\prod_{j\leq i}\prod_{J=1}^{N_{j}}\left( \lambda _{jJ}\right) ^{n_{jJ}^{iI}},
\label{generalbeta}
\end{equation}
where $f_{\{n_{jJ}^{iI}\}}\left( \alpha \right) $ are functions of the
marginal couplings of $\mathcal{C}$ and the sum is performed over the sets $%
\{n_{jJ}^{iI}\}$ of non-negative integers $n_{jJ}^{iI}$ such that

\begin{equation}
\sum_{j\leq i}j\sum_{J=1}^{N_{j}}n_{jJ}^{iI}=i.  \label{con2}
\end{equation}
The constant $\kappa $ is the only dimensionful parameter in the theory and
does not appear in the beta functions.

Due to (\ref{con2}), only a finite set of numbers $n_{jJ}^{iI}$ can be
greater than zero. This implies that the beta functions depend on the
irrelevant couplings in a polynomial way. Special sets $\{n_{jJ}^{iI}\}$
satisfying (\ref{con2}) are those where $n_{jJ}^{iI}$ is equal to one for $%
j=i$ and some index $J$, zero otherwise. It is useful to isolate this
contribution from the rest, obtaining
\begin{equation}
\beta _{iI}=\sum_{J=1}^{N_{i}}\gamma _{i}^{IJ}\left( \alpha \right) \lambda
_{iJ}+\delta _{iI},\qquad \qquad \delta
_{iI}=\sum_{\{m_{jJ}^{iI}\}}f_{\{m_{jJ}^{iI}\}}\left( \alpha \right)
\prod_{j<i}\prod_{J=1}^{N_{j}}\left( \lambda _{jJ}\right) ^{m_{jJ}^{iI}}.
\label{betage}
\end{equation}
Now the sum is performed over the sets $\{m_{jJ}^{iI}\}$ of non-negative
integers such that
\begin{equation}
\sum_{j<i}j\sum_{J=1}^{N_{j}}m_{jJ}^{iI}=i.  \label{**}
\end{equation}
The functions $\gamma _{i}^{IJ}\left( \alpha \right) $ are the entries of
the matrix $\gamma _{i}(\alpha )$ of anomalous dimensions of the operators $%
\mathcal{O}_{iI}$ of level $i$. The second term of (\ref{betage}) collects
the contributions of the operators $\mathcal{O}_{jJ}$ of levels $j<i$.
Observe that (\ref{**}) implies
\begin{equation}
\sum_{j<i}\sum_{J=1}^{N_{j}}m_{jJ}^{iI}\geq 2,  \label{***}
\end{equation}
which means that the beta function of $\lambda _{i}$ is at least quadratic
in the irrelevant couplings with $j<i$. \textit{A fortiori}, the $\delta
_{iI}$s vanish when all of the $\lambda _{iI}$s vanish. Indeed, at $\lambda
_{iI}=0$ the theory reduces to $\mathcal{L}_{\mathcal{C}}[\varphi ,\alpha ]$%
, which is finite by assumption.

\bigskip

\textbf{Deformation of level }$\ell $\textbf{. }Let $\gamma _{i}$ denote the
matrix having entries $\gamma _{i}^{IJ}\left( \alpha \right) $. The
deformation of level $\ell $ is defined as follows. First, set
\[
\lambda _{jJ}=0~~~\text{for ~}j\neq n\ell ,~n\text{ integer.}
\]
Using (\ref{betage}), this implies $\delta _{jJ}=\beta _{jJ}=0$ for~$j\neq
n\ell $ and $\delta _{\ell I}=0$. The equation
\[
\beta _{\ell I}=\frac{\mathrm{d}\lambda _{\ell I}}{\mathrm{d}\ln \mu }%
=\sum_{J=1}^{N_{\ell }}\gamma _{\ell }^{IJ}\left( \alpha \right) \lambda
_{\ell J}
\]
is solved by
\[
\lambda _{\ell I}(\mu )=\sum_{J=1}^{N_{\ell }}\exp \left( \gamma _{\ell }\ln
\mu /\overline{\mu }\right) ^{IJ}\lambda _{\ell J}(\overline{\mu }).
\]
The solution contains $N_{\ell }$ arbitrary parameters, which are the values
of the couplings $\lambda _{\ell I}$ at some reference scale $\overline{\mu }
$.

It is convenient to consider one arbitrary parameter at a time. The matrix $%
\gamma _{\ell }$ is real but in general its characteristic roots are
complex. For the moment I assume that $\gamma _{\ell }$ has at least one
real characteristic root, $r_{\ell }$, with multiplicity one. Let a tilde
denote vectors and matrices in a basis in which the matrix $\gamma _{\ell }$
has Jordan canonical form $\widetilde{\gamma }_{\ell }$ with $\widetilde{%
\gamma }_{\ell }^{11}=r_{\ell }$ (see for example \cite{lang}). Finally, let
$\widetilde{\lambda }_{\ell I}(\overline{\mu })=(\overline{\lambda }_{\ell
},0,\ldots 0)$. Then
\begin{equation}
\widetilde{\lambda }_{\ell }(\mu )=\exp \left( r_{\ell }\ln \mu /\overline{%
\mu }\right) \overline{\lambda }_{\ell }.  \label{run}
\end{equation}

For $n>1$, I write
\begin{equation}
\lambda _{n\ell I}=A_{n\ell I}\widetilde{\lambda }_{\ell }^{n}.  \label{la}
\end{equation}
The coefficients $A$ are scale invariant, i.e. invariant under the scale
symmetry (\ref{scala}). Quasi-finiteness is the requirement that the beta
functions of scale-invariant quantities vanish. If the $\ln \mu $
derivatives of both sides of (\ref{la}) are equated, the quasi-finiteness
equations read
\begin{equation}
\sum_{J=1}^{N_{\ell }}\widetilde{\gamma }_{n\ell }^{IJ}A_{n\ell J}^{{}}%
\widetilde{\lambda }_{\ell }^{n}+\delta _{n\ell I}=nr_{\ell }A_{n\ell I}%
\widetilde{\lambda }_{\ell }^{n}.  \label{system}
\end{equation}
This is a system of equations in the unknowns $A$. The solution can be
worked out inductively in $n$. Assuming that the systems (\ref{system}) have
been solved for $n=2,\ldots m-1$, and the solutions have the form (\ref{la}%
), then formula (\ref{betage}) ensures that the $\delta _{m\ell I}$s are
equal to known numbers times $\widetilde{\lambda }_{\ell }^{m}$. The $m^{th}$
system of equations (\ref{system}) can be solved if the matrix
\begin{equation}
\widehat{\gamma }_{m\ell }\equiv \gamma _{m\ell }-mr_{\ell }\mathbf{1}
\label{retta}
\end{equation}
is invertible, where \textbf{1} denotes the identity matrix. For real $%
r_{\ell }$ the invertibility of $\widehat{\gamma }_{m\ell }$ holds if and
only if no characteristic root of the matrix $\gamma _{m\ell }$ is equal to $%
m$ times $r_{\ell }$.

\bigskip

If the matrices $\widehat{\gamma }_{m\ell }$ are not invertible, solutions
exist if suitable entries of the vector $\delta _{m\ell }$ in (\ref{system})
vanish. In some cases a symmetry can ensure that certain irrelevant terms
have $\delta $ identically zero. Then the system (\ref{system}) can always
be solved. Operators with $\delta $ identically zero are called \textit{%
protected}. Examples of protected operators are the chiral operators in
four-dimensional supersymmetric theories \cite{grisaru}, which are discussed
in detail in section 6. At this stage of the discussion, it is convenient to
isolate the protected operators from the rest and concentrate the search for
solutions of the quasi-finiteness equations in the remaining subclass of
irrelevant terms. For simplicity, it is also convenient to set the couplings
of the protected operators to zero. Indeed, it is always possible to turn
those couplings on at a later stage. This operation is studied in section 6
and defines the protected irrelevant deformations. In the rest of this
section, I assume that the protected operators are dropped from (\ref{ola})
and that the $\lambda _{i}$s refer only to the unprotected irrelevant
operators, unless otherwise specified.

So, leaving the protected operators aside, the requirement that should be
satisfied for the existence of a consistent quasi-finite deformation of
level $\ell $, associated with the characteristic root $r_{\ell }$ of the
matrix $\gamma _{\ell }$, is the invertibility of the matrices (\ref{retta})
for $m>1$. Then the theory described by the lagrangian
\begin{equation}
\mathcal{L}[\varphi ]=\mathcal{L}_{\mathcal{C}}[\varphi ,\alpha ]+\widetilde{%
\kappa }~\widetilde{\mathcal{O}}_{\ell }(\varphi )+\sum_{n=1}^{\infty }%
\widetilde{\kappa }^{n}\sum_{I=1}^{N_{n\ell }}A_{n\ell I}(\alpha )~\mathcal{O%
}_{n\ell I}(\varphi ),  \label{finitesol}
\end{equation}
where $\widetilde{\kappa }=\kappa ^{\ell }\widetilde{\lambda }_{\ell }$, is
quasi-finite. The coefficients $A_{n\ell I}(\alpha )$ are uniquely specified
functions of the marginal couplings $\alpha $ of $\mathcal{C}$, determined
solving the system of equations (\ref{system}). The independent parameters
of (\ref{finitesol}) are $\alpha $ and $\widetilde{\kappa }$. The theory (%
\ref{finitesol}) is renormalized redefining the fields and the scale $%
\widetilde{\kappa }$, while the marginal couplings $\alpha $ and the
coefficients $A_{n\ell I}(\alpha )$ are unrenormalized. The scale $%
\widetilde{\kappa }$ is the unique parameter of the theory that
can run. The power-like divergences do not contribute to the RG
equations and so can be subtracted as they come, without adding
new independent couplings. The number $\ell $ is called
\textit{lowest level} of the deformation, the term
$\widetilde{\mathcal{O}}_{\ell }$ is the lowest-level operator of
the deformation and the sum in (\ref{finitesol}) is called
\textit{queue} of the deformation.

Now I now discuss the meaning of (\ref{retta}) and the existence of
solutions.

\bigskip

\textbf{Existence of solutions.} Neglecting, for pedagogical purposes, the
renormalization mixing for a moment, i.e. assuming that the indices $I,J$
can have only one value, the condition (\ref{retta}) reads
\begin{equation}
\gamma _{m\ell }\neq mr_{\ell }\text{,}  \label{dimma}
\end{equation}
for $m>1$, and says that the anomalous dimension of the irrelevant term of
level $m\ell $ should not be equal to $m$ times the anomalous dimension of
the lowest-level operator.

The meaning of the apparently obscure condition (\ref{dimma}) is actually
simple, as I now explain. The considerations that follow are not claimed to
be rigorous, but purely illustrative. Consider for concreteness a theory
containing scalar fields $\varphi $ in $d=4$ dimensions and restrict the
attention to operators without derivatives. Then
\begin{equation}
\mathrm{O}_{m\ell }\sim \varphi ^{d+m\ell }=\varphi ^{d}\left( \varphi
^{\ell }\right) ^{m},\qquad \mathrm{O}_{\ell }\sim \varphi ^{d+\ell
}=\varphi ^{d}\left( \varphi ^{\ell }\right) .  \label{inte}
\end{equation}
The operator $\mathrm{O}_{m\ell }$ is obtained sticking $m$ factors $\varphi
^{\ell }$ to the operator $\varphi ^{d}$. The operator $\varphi ^{d}$ has
level zero: it can be thought as a marginal deformation of $\mathcal{C}$ and
considered finite. Now, in renormalization theory, when operators are
multiplied together at the same point in spacetime, it is not sufficient to
renormalize the factors to renormalize the product, but it is necessary to
introduce a further renormalization constant for the product. The condition (%
\ref{dimma}) says in practice that the renormalization constant for the
product should be non-trivial. Common experience with renormalization
suggests that whenever a quantity can diverge (because it is not protected
by symmetries, power-counting, etc.), it generically does diverge. So,
excluding miraculous cancellations, it is reasonable to assume that the
products of operators have non-trivial renormalization constants and
therefore that the matrices (\ref{retta}) are invertible.

When this is true, it is possible to define the consistent irrelevant
deformations (\ref{finitesol}) of the conformal field theory $\mathcal{C}$.
Observe that, in any case, the invertibility or non-invertibility of the
matrices (\ref{retta}) is a property of the conformal theory $\mathcal{C}$,
so it is possible to say which irrelevant deformations are allowed from the
sole knowledge of $\mathcal{C}$, before actually deforming the theory.

There might exist situations in which the condition (\ref{dimma}) is valid
up to, say, $m=N$ and violated for $m>N$. Then, the number of parameters
necessary for the renormalization of this deformation remains constant up to
the order $\kappa ^{N\ell }$. At the order $\kappa ^{(N+1)\ell }$ the system
(\ref{system}) cannot be solved and (\ref{la}) cannot be imposed. This means
that new independent (running) parameters must be added at the level $%
(N+1)\ell $. This is a particular case of ``multiple'' deformation, in the
sense explained below. The deformation remains predictive at the level of
fundamental field theory if the total number of independent free parameters
remains finite.

\bigskip

\textbf{Multiple deformations.} It is possible to construct also
multiple deformations, of levels $\ell _{1},\cdots \ell _{k}$.
These deformations have more parameters that run independently.
For the moment, I\ still assume that the relevant characteristic
roots $r_{\ell j}$ are real with multiplicity one. Moreover, I
assume that the integers $\ell _{1},\cdots \ell _{k}$ are
relatively prime. Formula (\ref{la}) generalizes to
\begin{equation}
\lambda _{iI}=\sum_{\{n\}}A_{iI}^{n_{1}\cdots n_{k}}\widetilde{\lambda }%
_{\ell _{1}}^{n_{1}}\cdots \widetilde{\lambda }_{\ell _{k}}^{n_{k}},\qquad
\sum_{j=1}^{k}n_{j}\ell _{j}=i,\quad n_{j}\geq 0.  \label{ehi}
\end{equation}
The couplings that cannot be written in this form are set to zero. The
couplings $\widetilde{\lambda }_{n_{j}}$ run as
\[
\widetilde{\lambda }_{\ell _{j}}(\mu )=\exp \left( r_{\ell _{j}}\ln \mu /%
\overline{\mu }\right) \overline{\lambda }_{\ell _{j}}.
\]
Quasi-finiteness demands that the scale-invariant quantities $A$ have
vanishing beta functions. The system of equations (\ref{system}) generalizes
to
\begin{equation}
\sum_{J=1}^{N_{j}}\sum_{\{n\}}\gamma _{i}^{IJ}A_{iJ}^{n_{1}\cdots n_{k}}%
\widetilde{\lambda }_{\ell _{1}}^{n_{1}}\cdots \widetilde{\lambda }_{\ell
_{k}}^{n_{k}}+\delta _{iI}=\sum_{\{n\}}A_{iI}^{n_{1}\cdots n_{k}}\widetilde{%
\lambda }_{\ell _{1}}^{n_{1}}\cdots \widetilde{\lambda }_{\ell
_{k}}^{n_{k}}\sum_{j=1}^{k}n_{j}\overline{\gamma }_{\ell j}.  \label{compe}
\end{equation}
Proceeding inductively in the level $i$, formula (\ref{betagel}) shows that $%
\delta _{iI}$ can be expressed as
\[
\delta _{iI}=\sum_{\{n\}}\delta _{iI}^{n_{1}\cdots n_{k}}\widetilde{\lambda }%
_{\ell _{1}}^{n_{1}}\cdots \widetilde{\lambda }_{\ell _{k}}^{n_{k}},
\]
where $\delta _{iI}^{n_{1}\cdots n_{k}}$ are numbers determined solving the
systems (\ref{compe}) for levels $j<i$. The system (\ref{compe}) can be
split into a set of equations
\[
\sum_{J=1}^{N_{j}}\gamma _{i}^{IJ}A_{iJ}^{n_{1}\cdots
n_{k}}-A_{iI}^{n_{1}\cdots n_{k}}\sum_{j=1}^{k}n_{j}r_{\ell j}=-\delta
_{iI}^{n_{1}\cdots n_{k}}.
\]
It is immediate to see that the number of unknowns is equal to the number of
equations. There exists a unique solution if no characteristic root of the
matrix $\gamma _{i}$ is equal to
\begin{equation}
\sum_{j=1}^{k}n_{j}r_{\ell j},\qquad \text{with }n_{j}\text{ non-negative
integers such that }\sum_{j=1}^{k}n_{j}=i.  \label{rettadue}
\end{equation}
This requirement has an interpretation similar to the one of (\ref{inte}):
an operator of level $i$ can be written in many ways as the product of an
operator of level 0 and operators of positive levels $j<i$, but in no case
the anomalous dimension of the product should be equal to the sum of the
anomalous dimensions of the factors.

When the characteristic roots of the matrix $\gamma _{\ell }$ are complex or
have multiplicity greater than one, or when the levels $\ell _{1},\cdots
\ell _{k}$ are not relatively prime, the derivation generalizes
straightforwardly, as well as the requirement (\ref{rettadue}), but the
formulas become technically heavier. These generalizations do not teach
anything new and are left to the reader.

\bigskip

\textbf{Sufficient condition for the existence of a perturbative expansion}.
If $\mathcal{C}$ is a family of conformal field theories that becomes free
when some marginal coupling $\alpha $ tends to zero, then the irrelevant
deformation (\ref{finitesol}) might not admit a smooth $\alpha \rightarrow 0$
limit, due to the denominator appearing when the matrix (\ref{retta}) is
inverted. However, if the anomalous dimensions of the irrelevant couplings
satisfy a certain boundedness condition, it is possible to keep $\alpha $
small, but different from zero, and have a meaningful perturbative expansion
in powers of an effective $\kappa _{\mathrm{eff}}$.

Assume that there exist $\alpha $-independent numbers $c_{n}$ and a $\eta
>0~ $such that$~$%
\begin{equation}
\left| (\widehat{\gamma }_{n\ell }^{-1})^{IJ}\right| <\frac{c_{n}}{\eta }
\label{bound1}
\end{equation}
for every $n$. The quantity $\eta $ is a function of $\alpha $ (and $\ell $)
and tends to zero when $\alpha $ tends to zero. Then, it is possible to
prove that the solutions of (\ref{system}) behave not worse than
\begin{equation}
|A_{n\ell I}|\sim \widetilde{c}_{n}\frac{1}{\eta ^{(n-1)\ell }},
\label{ratio}
\end{equation}
where $\widetilde{c}_{n}$ are numbers and depend on the $c_{n}$s. The
behavior (\ref{ratio}) can be proved inductively in $n$. Indeed, if (\ref
{ratio}) is true for $n<m$, then (\ref{betage}), (\ref{system}) and (\ref
{***}) imply that (\ref{ratio}) is also true for $n=m$.

Under the assumption (\ref{bound1}), let us compare the behaviors of the
irrelevant terms of dimensionality $d+n\ell $ versus the behaviors of the
marginal terms of $\mathcal{C}$, as functions of the energy scale $E$ of a
physical process. The ratio between these two types of contributions behaves
not worse than
\[
a_{n}\eta ^{\ell }\left( \frac{\widetilde{\kappa }^{1/\ell }E}{\eta }\right)
^{n\ell },
\]
$a_{n}$ being calculable numbers that take care also of the $c_{n}$s. The
perturbative expansion in powers of the energy is meaningful for energies $E$
much smaller than the ``effective Planck mass''
\begin{equation}
\frac{1}{\kappa _{\mathrm{eff}}}\equiv \frac{\eta }{\widetilde{\kappa }%
^{1/\ell }}.  \label{keff}
\end{equation}
This up to the behavior of the numerical factors $a_{n}$, which cannot be
predicted before solving the theory.

\bigskip

In conclusion, to have consistent irrelevant deformations almost all of the
matrices (\ref{retta}) should be invertible and there must exist a $\eta >0$
satisfying (\ref{bound1}). ``Almost all'' means all but a finite number. The
restrictions concern only the renormalizable subsector $\mathcal{C}$ of the
theory and can be studied before turning the irrelevant deformation on. In
the free-field limit the effective Planck mass $1/\kappa _{\mathrm{eff}}$
tends to zero and the expansion in powers of the energy has zero convergence
radius. This is why the free field theories do not admit consistent
irrelevant deformations in the approach of this paper. This is also the
reason why the method of this paper and ref. \cite{mid} cannot be used to
quantize gravity in four dimensions.

\section{Pauli deformation of Yang-Mills theory coupled with fermions}

\setcounter{equation}{0}

In this and the next two sections, I apply the general approach of
the previous section to a concrete model, the Pauli deformation of
an interacting conformal field theory made of fermions and gauge
fields. I study the levels 1 and 2 of the deformation. The Pauli
term has dimensionality 5 and is multiplied by a running parameter
$\kappa $. In section 4 I study the self-renormalization of this
term. In section 5 I solve the quasi-finiteness equations and
compute the values of the couplings that multiply the irrelevant
terms of dimensionality 6 (four-fermion terms and $F^{3}$). I work
in the Euclidean framework and use the dimensional-regularization
technique.

\bigskip

The renormalized lagrangian of non-Abelian Yang-Mills theory coupled
massless fermions is
\begin{equation}
\mathcal{L}=\frac{\mu ^{-\varepsilon }}{4g^{2}Z_{g}^{2}}(\mathcal{F}_{\mu
\nu }^{a})^{2}+\overline{\Psi }_{i}^{I}\mathcal{D}\!\!\!\!\slash_{ij}\Psi
_{j}^{I},  \label{lagra}
\end{equation}
where $\mathcal{A}_{\mu }^{a}=Z_{A}^{1/2}A_{\mu }^{a}$, $\Psi
_{i}^{I}=Z_{\psi }^{1/2}\psi _{i}^{I}$ and $\mathcal{F}_{\mu \nu
}^{a}=\partial _{\mu }\mathcal{A}_{\nu }^{a}-\partial _{\nu }\mathcal{A}%
_{\mu }^{a}+f^{abc}\mathcal{A}_{\mu }^{b}\mathcal{A}_{\nu }^{c}$, $\mathcal{D%
}_{\mu }^{ij}\Psi _{j}^{I}=\partial _{\mu }\Psi _{i}^{I}+\mathcal{A}_{\mu
}^{a}T_{ij}^{a}\Psi _{j}^{I}$, The index $I=1,\ldots N_{f}$ is a flavor
index. I assume that the fermions are in the fundamental representation.
Details about the notation are given in the appendix.

The gauge-fixing part reads
\begin{equation}
\mathcal{L}_{gf}=\frac{\mu ^{-\varepsilon }}{2\alpha g^{2}}(\partial _{\mu
}A_{\mu }^{a})^{2}+Z_{C}\overline{C}^{a}\partial _{\mu }\mathcal{D}_{\mu
}^{ab}C^{b},  \label{gf}
\end{equation}
where $\mathcal{D}_{\mu }^{ab}C^{b}=\partial _{\mu }C^{a}+f^{abc}\mathcal{A}%
_{\mu }^{b}C^{c}$.

The renormalization constants are, to the first loop order (see for example
\cite{muta})
\begin{eqnarray}
Z_{\psi } &=&1-\frac{g^{2}\alpha }{8\pi ^{2}\varepsilon }\frac{N_{c}^{2}-1}{%
2N_{c}}\equiv 1+\delta Z_{\psi },\qquad \qquad Z_{A}=1-\frac{g^{2}N_{c}}{%
16\pi ^{2}\varepsilon }\left( 3+\alpha \right) \equiv 1+\delta Z_{A},
\nonumber \\
Z_{C} &=&1+\frac{g^{2}N_{c}}{32\pi ^{2}\varepsilon }(3-\alpha ),\qquad
\qquad \qquad \qquad \quad Z_{g}=1-\frac{g^{2}}{48\pi ^{2}\varepsilon }%
\left( 11N_{c}-2N_{f}\right) .  \label{wvrnc}
\end{eqnarray}

\bigskip

\textbf{IR interacting fixed point.} To identify the IR\ fixed point, it is
necessary to write the beta function of $g$ to the second loop order. In the
limit where $N_{c}$ and $N_{f}$ are large, $g$ is small, but $g^{2}N_{c}$
and $N_{f}/N_{c}$ are fixed, and such that $N_{f}/N_{c}\lesssim 11/2$ , the
beta function reads \cite{muta}
\begin{equation}
\frac{\beta _{g}}{g}=-\frac{\Delta }{3}\frac{g^{2}N_{c}}{16\pi ^{2}}+\frac{25%
}{2}\left( \frac{g^{2}N_{c}}{16\pi ^{2}}\right) ^{2}+\sum_{n=3}^{\infty
}c_{n}\left( \frac{g^{2}N_{c}}{16\pi ^{2}}\right) ^{n},  \label{beta2}
\end{equation}
where $\Delta \equiv 11-2N_{f}/N_{c}\ll 1$ and the $c_{n}$s are numerical
coefficients. In the limit just described, the theory is asymptotically free
and the first two contributions of the beta function have opposite signs.
Moreover, the first contribution is arbitrarily small. This ensures that,
expanding in powers of $\Delta $, the beta function has a second zero for
\begin{equation}
\frac{g_{*}^{2}N_{c}}{16\pi ^{2}}=\frac{2}{75}\Delta +\mathcal{O}(\Delta
^{2}).  \label{zero}
\end{equation}
This zero defines a non-trivial conformal field theory.

The purpose of this section and sections 4-5 is to construct the Pauli
deformation of this interacting conformal field theory. I concentrate on the
irrelevant terms of levels $1$ and $2$.

\bigskip

\textbf{Irrelevant terms of level }$\mathbf{1}$\textbf{.} There is only one
irrelevant term of level $1$, the Pauli term
\begin{equation}
\mathcal{L}_{\mathrm{Pauli}}=\kappa \lambda Z_{\lambda }~\mathcal{F}_{\mu
\nu }^{a}~\overline{\Psi }_{i}^{I}T_{ij}^{a}\sigma _{\mu \nu }\Psi _{j}^{I},
\label{pau}
\end{equation}
where $\sigma _{\mu \nu }=-i[\gamma _{\mu },\gamma _{\nu }]/2$. The terms
\[
\kappa \overline{\Psi }\mathcal{D}\!\!\!\!\slash\mathcal{D}\!\!\!\!\slash%
\Psi ,\qquad \kappa \overline{\Psi }\mathcal{D}^{2}\Psi
\]
can be converted to (\ref{pau}) up to $\mathcal{O}(\kappa ^{2})$ using the
field equations (see the appendix).

\bigskip

\textbf{Irrelevant terms of level }$\mathbf{2}$\textbf{.} The classification
of the irrelevant terms of level $2$, instead, is more involved. First,
there is a unique term of level $2$ that does not contain fermion fields.
This is the $F^{3}$-term
\begin{equation}
\mathcal{L}_{F^{3}}=\frac{\kappa ^{2}\mu ^{-\varepsilon }}{6!}\zeta Z_{\zeta
}~f^{abc}\mathcal{F}_{\mu \nu }^{a}\mathcal{F}_{\nu \rho }^{b}\mathcal{F}%
_{\rho \mu }^{c}.  \label{fter}
\end{equation}
Other terms that do not contain fermions, such as
\[
\kappa ^{2}\mathcal{F}_{\mu \nu }^{a}\mathcal{D}^{2}\mathcal{F}_{\mu \nu
}^{a},\text{\qquad }\kappa ^{2}(\mathcal{D}_{\rho }\mathcal{F}_{\rho \mu
}^{a})^{2},
\]
can be converted to (\ref{fter}) plus four-fermion terms using the Bianchi
identity and the field equations, up to $\mathcal{O}(\kappa ^{3})$.

The independent four-fermion vertices are ten, precisely
\begin{eqnarray}
\mathrm{S} &=&(\overline{\Psi }_{i}^{I}\Psi _{i}^{I})^{2},\qquad \mathrm{P}=(%
\overline{\Psi }_{i}^{I}\gamma _{5}\Psi _{i}^{I})^{2},\qquad \mathrm{V}=(%
\overline{\Psi }_{i}^{I}\gamma _{\mu }\Psi _{i}^{I})^{2},\qquad \mathrm{A}=(%
\overline{\Psi }_{i}^{I}\gamma _{5}\gamma _{\mu }\Psi _{i}^{I})^{2},
\nonumber \\
\mathrm{T} &=&(\overline{\Psi }_{i}^{I}\sigma _{\mu \nu }\Psi
_{i}^{I})^{2},\qquad \mathrm{S}^{\prime }=(\overline{\Psi }_{i}^{I}\Psi
_{j}^{I})(\overline{\Psi }_{j}^{I}\Psi _{i}^{I}),\qquad \mathrm{P}^{\prime
}=(\overline{\Psi }_{i}^{I}\gamma _{5}\Psi _{j}^{I})(\overline{\Psi }%
_{j}^{I}\gamma _{5}\Psi _{i}^{I}),  \nonumber \\
\mathrm{V}^{\prime } &=&(\overline{\Psi }_{i}^{I}\gamma _{\mu }\Psi
_{i}^{I})^{2},\qquad \mathrm{A}^{\prime }=(\overline{\Psi }_{i}^{I}\gamma
_{5}\gamma _{\mu }\Psi _{i}^{I})^{2},\qquad \mathrm{T}^{\prime }=(\overline{%
\Psi }_{i}^{I}\sigma _{\mu \nu }\Psi _{j}^{I})(\overline{\Psi }%
_{j}^{I}\sigma _{\mu \nu }\Psi _{i}^{I}).  \label{basis}
\end{eqnarray}
The proof that these ten vertices are a basis is done using Fierz
identities. Every fermion has three indices: Lorentz, gauge and flavor. We
have to study the contractions of
\[
\overline{\Psi }_{i}^{I\alpha }\Psi _{j}^{I\beta }\overline{\Psi }%
_{k}^{J\gamma }\Psi _{l}^{J\delta }.
\]
The Clifford algebra contains 5 elements $\Gamma ^{A}$ (scalar,
pseudoscalar, vector, pseudovector and tensor). Parity and Lorentz
invariance impose that only the pairings $\Gamma ^{A}\times \Gamma ^{A}$
(i.e. 1$\times $1, $\gamma _{5}\times \gamma _{5}$, $\gamma _{\mu }\times
\gamma _{\mu }$, etc.) are allowed. The matrices $\Gamma ^{A}$ can contract
the Lorentz indices in two ways ($\alpha $ and $\gamma $ or $\alpha $ and $%
\delta $), but Fierz identities relate these two contractions. Consequently,
the Lorentz indices can be contracted in 5 independent ways. Finally, the
gauge indices can be contracted in two ways: $i$ with $j$ or $i$ with $l$.
So, in total there exist 10 independent contractions, the ones of (\ref
{basis}). This proves the statement.

The four-fermion lagrangian is written as
\begin{eqnarray}
\mathcal{L}_{\mathrm{4}\text{\textrm{F}}} &=&\frac{\kappa ^{2}\mu
^{\varepsilon }}{4}\left[ \xi _{1}Z_{\xi _{1}}~\mathrm{S}+\xi _{2}Z_{\xi
_{2}}~\mathrm{P}+\xi _{3}Z_{\xi _{3}}~\mathrm{S}^{\prime }\right. +\xi
_{4}Z_{\xi _{4}}~\mathrm{P}^{\prime }+\lambda _{1}Z_{\lambda _{1}}~\mathrm{V}%
+\lambda _{2}Z_{\lambda _{2}}~\mathrm{A}+  \nonumber \\
&&+\left. \lambda _{3}Z_{\lambda _{3}}~\mathrm{V}^{\prime }+\lambda
_{4}Z_{\lambda _{4}}~\mathrm{A}^{\prime }+\eta _{1}Z_{\eta _{1}}~\mathrm{T}%
+\eta _{2}Z_{\eta _{2}}~\mathrm{T}^{\prime }\right] .  \label{4f}
\end{eqnarray}

There exist no other independent parity-invariant lagrangian terms of level $%
2$. The terms containing two fermions and gauge fields, such as
\[
\mathcal{F}_{\mu \nu }^{a}~(\mathcal{D}_{\mu }\overline{\Psi }%
^{I})T^{a}\gamma _{\nu }^{I}\Psi ,\qquad \qquad \varepsilon _{\mu \nu \rho
\sigma }\mathcal{F}_{\mu \nu }^{a}~(\mathcal{D}_{\mu }\overline{\Psi }%
^{I})T^{a}\gamma _{\nu }\gamma _{5}^{I}\Psi ,
\]
are not independent. Using the field equations and Bianchi identities, they
can be converted into four-fermion terms, up to total derivatives and $%
\mathcal{O}(\kappa ^{3})$. The proof is given in the appendix, see formulas (%
\ref{uno}) and (\ref{due}).

\bigskip

\textbf{Pauli deformation.} The Pauli deformation of the theory (\ref{lagra}%
) is described by the lagrangian
\begin{equation}
\mathcal{L}=\frac{\mu ^{-\varepsilon }}{4g^{2}Z_{g}^{2}}(\mathcal{F}_{\mu
\nu }^{a})^{2}+\overline{\Psi }_{i}^{I}D\!\!\!\!\slash_{ij}\Psi _{j}^{I}+%
\mathcal{L}_{\mathrm{Pauli}}+\mathcal{L}_{F^{3}}+\mathcal{L}_{\mathrm{4}%
\text{\textrm{F}}}+\mathcal{O}(\kappa ^{3}).  \label{paulidef}
\end{equation}
The couplings of levels $>1$ have to be determined iteratively as explained
in section 2. This is illustrated in section 5 for level 2.

The field redefinitions have the form $\mathcal{A}_{\mu
}^{a}=Z_{A}^{1/2}A_{\mu }^{a}+\mathcal{O}(\kappa )$, $\Psi _{i}^{I}=Z_{\psi
}^{1/2}\psi _{i}^{I}+\mathcal{O}(\kappa )$. The Pauli vertex is the
lowest-level operator of the deformation (\ref{paulidef}). The four-fermion
vertices, the $F^{3}$ term and the $\mathcal{O}(\kappa ^{3})$-terms are the
queue of the deformation.

In the next two sections I perform a complete one-loop calculation up to $%
\mathcal{O}(\kappa ^{3})$ excluded. The calculation can be divided in two
steps. The first step is the renormalization of the Pauli coupling (section
4). This determines the coherent running of the Pauli deformation (\ref
{paulidef}), including the queue. At this level it is necessary to work out
also the field redefinitions explicitly, because they can be important for
the $\mathcal{O}(\kappa ^{2})$-calculations. The second step (section 5) is
the renormalization of the first terms of the queue (level 2), that is to
say the $F^{3}$-vertex and the four-fermion vertices. This calculation can
be divided itself into two parts, the self-renormalization of the level-2
vertices and their generation from two Pauli insertions. Using gauge
invariance it is possible to reduce the number of diagrams. Counterterms
with external legs $\overline{\psi }$-$\psi $, $A$-$\overline{\psi }$-$\psi $%
, $\overline{\psi }$-$\psi $-$\overline{\psi }$-$\psi $, $A$-$A$ and $A$-$A$-%
$A$ need to be calculated explicitly, but counterterms of the form $A$-$A$-$%
\overline{\psi }$-$\psi $ and $A$-$A$-$A$-$A$, for example, are related by
gauge invariance to the previous ones.

\section{Renormalization of the Pauli coupling}

\setcounter{equation}{0}

\begin{figure}[tbp]
\centerline{\epsfig{figure=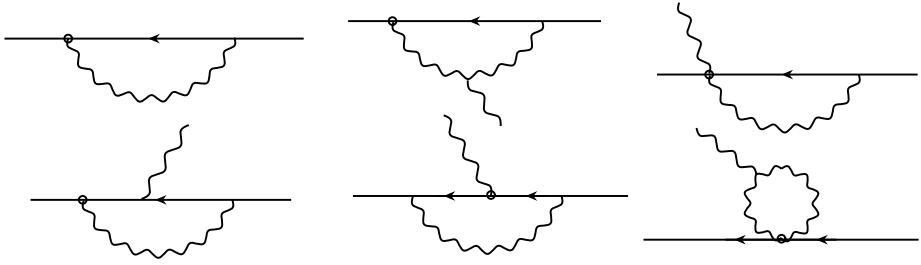,height=4cm,width=13cm}}
\caption{Self-renormalization of the Pauli term}
\label{fig3}
\end{figure}

The non-trivial graphs containing one insertion of the Pauli vertex are
shown in Fig. \ref{fig3}. Using the identity (\ref{one}), the associated
counterterms are
\begin{eqnarray}
\Delta \mathcal{L}_{\mathrm{Pauli}} &=&\frac{3ig^{2}\kappa \lambda }{8\pi
^{2}\varepsilon }\frac{N_{c}^{2}-1}{N_{c}}\overline{\psi }D\!\!\!\!\slash%
^{2}\psi +\frac{g^{2}\kappa \lambda }{64\pi ^{2}\varepsilon N_{c}}\left[
N_{c}^{2}(1-6\alpha )+4(\alpha -5)\right] \overline{\psi }T^{a}\sigma _{\mu
\nu }\psi ~F_{\mu \nu }^{a}+  \nonumber \\
&&+\frac{3ig^{2}\kappa \lambda N_{c}}{32\pi ^{2}\varepsilon }\left(
\overline{\psi }T^{a}\psi ~\partial _{\mu }A_{\mu }^{a}+2\partial _{\mu }%
\overline{\psi }T^{a}\psi ~A_{\mu }^{a}\right) +\mathcal{O}(\overline{\psi }%
A^{2}\psi ).  \label{deltalpauli}
\end{eqnarray}
Using (\ref{olb}), the terms proportional to the $\mathcal{O}(\kappa ^{0})$%
-field equations can be rabsorbed by means of the field redefinitions
\begin{eqnarray}
\Psi _{i}^{I} &=&Z_{\psi }^{1/2}\psi _{i}^{I}+\frac{3ig^{2}\kappa \lambda }{%
16\pi ^{2}\varepsilon }\left( \frac{N_{c}^{2}-1}{N_{c}}D\!\!\!\!\slash%
_{ij}\psi _{j}^{I}-\frac{N_{c}}{2}A\!\!\!\!\slash^{a}T_{ij}^{a}\psi _{j}^{I}+%
\mathcal{O}(A^{2}\psi )\right) +\mathcal{O}(\kappa ^{2}),  \nonumber \\
\mathcal{A}_{\mu }^{a} &=&Z_{A}^{1/2}A_{\mu }^{a}+\mathcal{O}(\kappa ^{2})%
\text{.}  \label{fref}
\end{eqnarray}
Observe that the fermion field redefinition is non-covariant. Isolating the
contributions of these field redefinitions inside (\ref{deltalpauli}) the
remaining counterterms are
\[
\frac{g^{2}\kappa \lambda }{32\pi ^{2}N_{c}\varepsilon }\left[
-N_{c}^{2}(1+3\alpha )+2(\alpha -5)\right] \overline{\psi }T^{a}\sigma _{\mu
\nu }\psi ~F_{\mu \nu }^{a}.
\]
The dependence on the gauge-fixing parameter $\alpha $ drops out factorizing
$Z_{\psi }Z_{A}^{1/2}$ in front of the Pauli term. Finally, the net
renormalization constant of the Pauli coupling $\lambda $ is
\[
Z_{\lambda }=1+\frac{g^{2}(N_{c}^{2}-5)}{16\pi ^{2}N_{c}\varepsilon }+%
\mathcal{O}(g^{4}).
\]
This gives the beta function
\[
\beta _{\lambda }=\frac{g^{2}\lambda (N_{c}^{2}-5)}{16\pi ^{2}N_{c}}\sim
\frac{2}{75}\lambda \Delta
\]
and the running behavior
\[
\lambda (\Lambda )=\lambda (\mu )\left( \frac{\Lambda }{\mu }\right)
^{2\Delta /75}.
\]
The Pauli coupling is IR free, but this fact is not necessary for the
consistency of the perturbative expansion. The reason is that perturbation
theory can only generate logarithmic corrections, while the irrelevant
deformations contain powers of the energy. At energy $E$ the behavior of the
Pauli term versus the behavior of, say, the term $F^{2}$ is
\begin{equation}
\sim \lambda (\mu )\left( \frac{E}{\mu }\right) ^{2\Delta /75}(\kappa E).
\label{beha}
\end{equation}
When the energy is small with respect to $1/\kappa $ and $\mu $ (it is
possible to choose $\mu \sim 1/\kappa $ without loss of generality) the
behavior (\ref{beha}) is compatible with the perturbative expansion in
powers of the energy if $\Delta $ is greater than $-75/2$.

Observe that the gauge field has no $\mathcal{O}(\kappa )$-field
redefinition. This is good, because it ensures that the gauge-fixing sector
of the theory is unmodified to this order. It is easy to check that no
one-loop divergent graph with external ghost legs and one Pauli vertex can
be constructed.

Moreover, the form of the $\mathcal{O}(\kappa )$ corrections to the field
redefinitions, shown in (\ref{fref}), ensures that these corrections can be
ignored to order $\mathcal{O}(\kappa ^{2})$, because they do not contribute
to the renormalization of the essential couplings. Indeed, they produce $%
\mathcal{O}(\kappa ^{2})$-contributions that are either proportional to the $%
\mathcal{O}(\kappa ^{0})$ fermion field equations or have the form $A$-$A$-$%
\overline{\psi }$-$\psi $.

\section{The Pauli deformation to order $\mathcal{O}(\kappa ^{2})$}

\setcounter{equation}{0}
\begin{figure}[tbp]
\centerline{\epsfig{figure=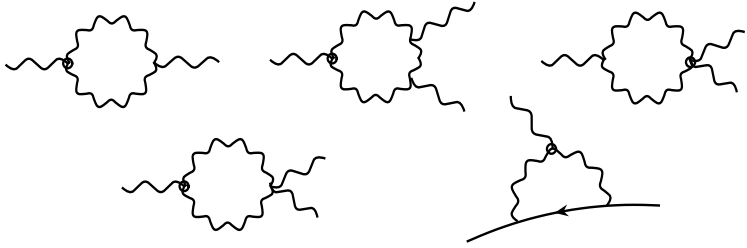,height=5cm,width=13cm}}
\caption{Renormalization of the $F^{3}$ vertex}
\label{fig7}
\end{figure}
In this section I study the level-2 terms of the queue of the Pauli
deformation. First I compute the relevant Feynman diagrams and then solve
the quasi-finiteness equations that determine the values of the couplings
multiplying the $F^{3}$ term and the four-fermion vertices of (\ref{paulidef}%
).

\bigskip

\textbf{Renormalization of the }$F^{3}$\textbf{\ term. }The diagrams
containing one insertion of the $F^{3}$-vertex are depicted in Fig. \ref
{fig7}. The counterterms for these graphs are
\begin{eqnarray}
\Delta \mathcal{L}_{F^{3}} &=&-\frac{g^{2}\kappa ^{2}N_{c}\zeta \mu
^{-\varepsilon }}{16\pi ^{2}\varepsilon }\left( D_{\mu }^{ab}F_{\mu \nu
}^{b}\right) ^{2}+\frac{3(5-\alpha )g^{2}N_{c}}{32\pi ^{2}\varepsilon }\frac{%
\zeta \kappa ^{2}\mu ^{-\varepsilon }}{6!}~f^{abc}F_{\mu \nu }^{a}F_{\nu
\rho }^{b}F_{\rho \mu }^{c}  \nonumber \\
&&+\frac{g^{4}\kappa ^{2}N_{c}\zeta }{16\pi ^{2}\varepsilon }\overline{\psi }%
T^{a}\gamma _{\nu }\psi ~D_{\mu }^{ab}F_{\mu \nu }^{b}  \label{deltalf3}
\end{eqnarray}
plus terms proportional to $\partial _{\mu }A_{\mu }^{a}$ (which can be
subtracted with a redefinition of the gauge fixing), total derivatives,
terms of the form $A$-$A$-$A$-$A$ and $A$-$A$-$\overline{\psi }$-$\psi $ and
terms proportional to the field equations (\ref{field2}).

Using the field equations, the first and third terms of (\ref{deltalf3})
mutually cancel. Finally, isolating the contribution
\[
\frac{3}{2}\delta Z_{A}\frac{\zeta \kappa ^{2}\mu ^{-\varepsilon }}{6!}%
~f^{abc}F_{\mu \nu }^{a}F_{\nu \rho }^{b}F_{\rho \mu }^{c}
\]
of the wave-function renormalization constants, the $\alpha $-dependence
drops out, as it should be, and the net result contributing to $Z_{\zeta }$
is
\begin{equation}
\Delta \mathcal{L}_{F^{3}\text{-net}}=\frac{3g^{2}N_{c}}{4\pi
^{2}\varepsilon }\frac{\zeta \kappa ^{2}\mu ^{-\varepsilon }}{6!}%
~f^{abc}F_{\mu \nu }^{a}F_{\nu \rho }^{b}F_{\rho \mu }^{c}.
\label{netdeltalf3}
\end{equation}
The renormalization constant $Z_{\zeta }$ receives contributions also from
graphs containing two insertions of the Pauli vertex (see below).
\begin{figure}[tbp]
\centerline{\epsfig{figure=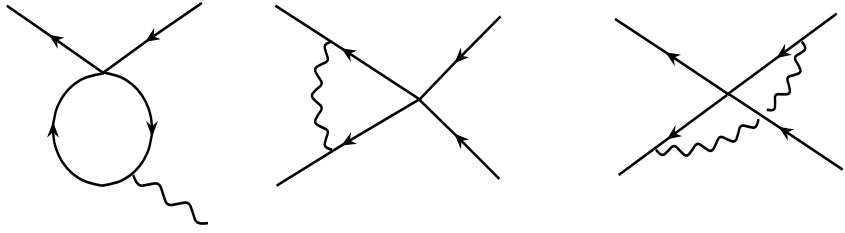,height=4cm,width=13cm}}
\caption{Self-renormalization of the four-fermion vertices}
\label{fig1}
\end{figure}

\bigskip

\textbf{Renormalization of the four-fermion terms.} The divergent graphs
containing a four-fermion vertex are shown in Fig. \ref{fig1}. The
counterterms
\begin{equation}
\Delta \mathcal{L}_{\mathrm{4}\text{\textrm{F}}}=2\delta Z_{\psi }Z_{\psi
}^{-2}Z_{\zeta }^{-1}\mathcal{L}_{\mathrm{4}\text{\textrm{F}}}+\Delta
\mathcal{L}_{\mathrm{vf}}+\Delta \mathcal{L}_{\mathrm{4}\text{\textrm{f}}},
\label{deltal4f}
\end{equation}
can be split in three parts: the contributions associated with the
wave-function renormalization constant $Z_{\psi }$, which reabsorb every
gauge-fixing dependence, the vertex counterterm
\begin{equation}
\Delta \mathcal{L}_{\mathrm{vf}}=-\frac{\kappa ^{2}}{48\pi ^{2}\varepsilon }%
\left( 2\lambda _{1}+2\lambda _{2}+4N_{f}\lambda _{3}-\xi _{1}+\xi
_{2}\right) \overline{\psi }T^{a}\gamma _{\nu }\psi ~D_{\mu }^{ab}F_{\mu \nu
}^{b}  \label{lv4f1}
\end{equation}
and the four-fermion counterterms
\begin{eqnarray}
\Delta \mathcal{L}_{\mathrm{4}\text{\textrm{f}}} &=&-\frac{g^{2}}{32\pi
^{2}\varepsilon N_{c}}\frac{\kappa ^{2}\mu ^{\varepsilon }}{4}\left[
12\left( \xi _{1}(N_{c}^{2}-1)+2\eta _{2}N_{c}+\xi _{3}N_{c}-4\eta
_{1}\right) \mathrm{S}+\right.  \nonumber \\
&&+12\left( \xi _{2}(N_{c}^{2}-1)+2\eta _{2}N_{c}+\xi _{4}N_{c}-4\eta
_{1}\right) \mathrm{P}+12\left( 2\eta _{2}(N_{c}^{2}-2)+4\eta _{1}N_{c}-\xi
_{3}\right) \mathrm{S}^{\prime }+  \nonumber \\
&&+12\left( 2\eta _{2}(N_{c}^{2}-2)+4\eta _{1}N_{c}-\xi _{4}\right) \mathrm{P%
}^{\prime }+6\left( 2\lambda _{2}-\lambda _{3}N_{c}-\lambda _{4}N_{c}\right)
\mathrm{V}+  \nonumber \\
&&+6\left( 2\lambda _{1}-\lambda _{3}N_{c}-\lambda _{4}N_{c}\right) \mathrm{A%
}+6\left( \lambda _{3}N_{c}^{2}-2\lambda _{2}N_{c}-\lambda
_{4}(N_{c}^{2}-2)\right) \mathrm{V}^{\prime }+  \nonumber \\
&&+6\left( \lambda _{4}N_{c}^{2}-2\lambda _{1}N_{c}-\lambda
_{3}(N_{c}^{2}-2)\right) \mathrm{A}^{\prime }+  \nonumber \\
&&+\left( (\xi _{3}+\xi _{4})N_{c}-2(\xi _{1}+\xi _{2})-12\eta
_{2}N_{c}-4\eta _{1}(N_{c}^{2}-1)\right) \mathrm{T}+  \nonumber \\
&&\left. +\left( (\xi _{3}+\xi _{4})(N_{c}^{2}-2)+2N_{c}(\xi _{1}+\xi
_{2})+4\eta _{2}(2N_{c}^{2}+1)\right) \mathrm{T}^{\prime }\right] .
\label{l4f}
\end{eqnarray}

Using the field equations and the identity (\ref{id1}), the vertex
counterterm (\ref{lv4f1}) is converted into the sum of two four-fermion
terms:
\begin{equation}
\Delta \mathcal{L}_{\mathrm{vf}}\rightarrow -\frac{g^{2}\kappa ^{2}\mu
^{\varepsilon }}{96\pi ^{2}\varepsilon N_{c}}\left( 2\lambda _{1}+2\lambda
_{2}+4N_{f}\lambda _{3}-\xi _{1}+\xi _{2}\right) \left( \mathrm{V}-N_{c}%
\mathrm{V}^{\prime }\right) .  \label{lv4f}
\end{equation}

\bigskip

\begin{figure}[tbp]
\centerline{\epsfig{figure=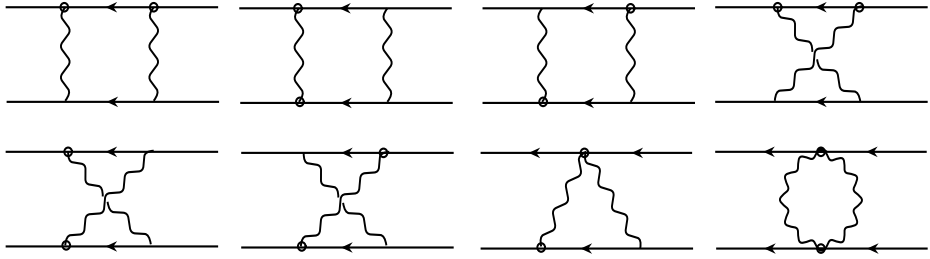,height=4cm,width=13cm}}
\caption{Four-fermion renormalization to order $\lambda ^{2}$}
\label{fig2}
\end{figure}

\begin{figure}[tbp]
\centerline{\epsfig{figure=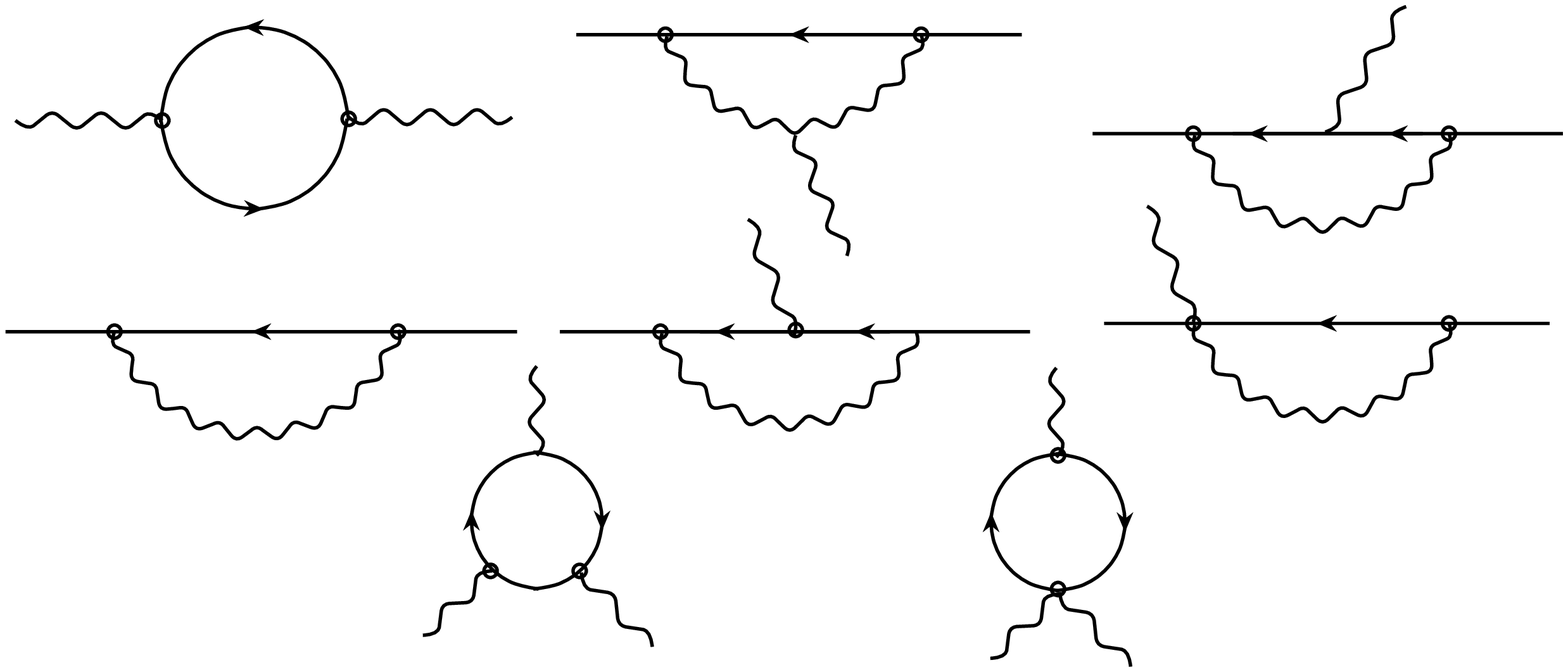,height=6cm,width=13cm}}
\caption{Renormalization of two- and three-point functions to order $\lambda
^{2}$}
\label{fig4}
\end{figure}

\textbf{Generation of }$F^{3}$\textbf{\ and four-fermion terms from two
Pauli insertions. }There remain to study the contributions of type $\delta $
in (\ref{betagel}). These are due to the graphs containing two insertions of
Pauli vertices. The graphs are grouped into two sets, depicted in Figs. \ref
{fig2} and \ref{fig4}.

The contributions of the graphs of Fig. \ref{fig2} are
\begin{eqnarray}
\Delta \mathcal{L}_{\mathrm{4}\text{\textrm{F-}}\lambda ^{2}}=\frac{%
g^{4}\lambda ^{2}}{8\pi ^{2}\varepsilon N_{c}^{2}}\frac{\kappa ^{2}\mu
^{\varepsilon }}{4} &&\left[ -24(N_{c}^{2}+2)\mathrm{S}\right.
-24N_{c}(N_{c}^{2}-4)\mathrm{S}^{\prime }+3N_{c}^{2}\mathrm{V}+6(N_{c}^{2}+2)%
\mathrm{A}+  \nonumber \\
&&-3N_{c}^{3}\mathrm{V}^{\prime }+6N_{c}(N_{c}^{2}-4)\mathrm{A}^{\prime
}+N_{c}^{2}\mathrm{T}\left. -N_{c}^{3}\mathrm{T}^{\prime }\right] .
\label{l4fl2}
\end{eqnarray}

The contributions of the graphs of Fig. \ref{fig4} are
\begin{eqnarray}
\Delta \mathcal{L}_{F^{3}\text{-}\lambda ^{2}} &=&-\frac{\lambda
^{2}N_{f}\kappa ^{2}\mu ^{-\varepsilon }}{12\pi ^{2}\varepsilon }\left(
D_{\mu }^{ab}F_{\mu \nu }^{b}\right) ^{2}-\frac{\lambda ^{2}N_{f}}{4\pi
^{2}\varepsilon }\frac{\kappa ^{2}\mu ^{-\varepsilon }}{6!}~f^{abc}F_{\mu
\nu }^{a}F_{\nu \rho }^{b}F_{\rho \mu }^{c}+  \nonumber \\
&&+\frac{g^{2}\lambda ^{2}\kappa ^{2}}{48\pi ^{2}\varepsilon N_{c}}\left[
(23N_{c}^{2}+10)\overline{\psi }T^{a}\gamma _{\nu }\psi ~D_{\mu }^{ab}F_{\mu
\nu }^{b}+12(N_{c}^{2}-1)F_{\mu \nu }^{a}~D_{\mu }\overline{\psi }%
T^{a}\gamma _{\nu }\psi \right. +  \nonumber \\
&&+\left. 12N_{c}^{2}~\varepsilon _{\mu \nu \rho \sigma }F_{\mu \nu
}^{a}~D_{\rho }\overline{\psi }T^{a}\gamma _{\sigma }\gamma _{5}\psi \right]
\label{lf31}
\end{eqnarray}
plus terms proportional to $\partial _{\mu }A_{\mu }^{a}$, total
derivatives, terms of the form $A$-$A$-$A$-$A$ and $A$-$A$-$\overline{\psi }$%
-$\psi $ and terms proportional to the field equations. Using the identities
(\ref{uno}) and (\ref{due}) and the field equations, the counterterm (\ref
{lf31}) can be re-written as
\begin{equation}
\Delta \mathcal{L}_{F^{3}\text{-}\lambda ^{2}}\rightarrow \frac{\lambda
^{2}g^{4}\kappa ^{2}\mu ^{\varepsilon }}{96\pi ^{2}\varepsilon N_{c}^{2}}%
\left( 5N_{c}^{2}+16-4N_{f}N_{c}\right) \left( \mathrm{V}-N_{c}\mathrm{V}%
^{\prime }\right) -\frac{\lambda ^{2}N_{f}}{4\pi ^{2}\varepsilon }\frac{%
\kappa ^{2}\mu ^{-\varepsilon }}{6!}~f^{abc}F_{\mu \nu }^{a}F_{\nu \rho
}^{b}F_{\rho \mu }^{c}.  \label{lf3l2}
\end{equation}

\bigskip

\textbf{The Pauli deformation to order }$\mathcal{O}(\kappa ^{2})$.
Recapitulating, the $\mathcal{O}(\kappa ^{2})$ counterterms, cleaned of the
contributions due to the wave-function renormalization constants and the
terms proportional to the gauge-fixing and the field equations, are the sum
of (\ref{netdeltalf3}), (\ref{l4f}), (\ref{lv4f}), (\ref{l4fl2}) and (\ref
{lf3l2}):
\[
\Delta \mathcal{L}_{\text{net}}=\Delta \mathcal{L}_{F^{3}\text{-net}}+\Delta
\mathcal{L}_{\mathrm{4}\text{\textrm{f}}}+\Delta \mathcal{L}_{\mathrm{vf}%
}+\Delta \mathcal{L}_{\mathrm{4}\text{\textrm{F-}}\lambda ^{2}}+\Delta
\mathcal{L}_{F^{3}\text{-}\lambda ^{2}}.
\]

It is convenient to start from the $F^{3}$-terms, which can be easily
isolated from the rest. The $\zeta $-renormalization constant is
\[
\zeta Z_{\zeta }=\zeta \left( 1+\frac{3g^{2}N_{c}}{4\pi ^{2}\varepsilon }%
\right) ~-\frac{\lambda ^{2}N_{f}}{4\pi ^{2}\varepsilon }.
\]
The quasi-finiteness equations relate $\zeta $ to $\lambda $ in such a way
that the scale-invariant combination
\[
u\equiv \frac{\zeta }{\lambda ^{2}}
\]
has vanishing beta function. This means also
\[
\zeta Z_{\zeta }=\zeta Z_{\lambda }^{2}\text{.}
\]
The result is
\begin{equation}
\zeta ~=\frac{11\lambda ^{2}}{5g_{*}^{2}}=\frac{165}{2}\frac{1}{\Delta }%
\left( \frac{\lambda ^{2}N_{c}}{16\pi ^{2}}\right) .  \label{zita}
\end{equation}
The beta function of $\zeta $ is
\[
\beta _{\zeta }=\frac{3g_{*}^{2}\zeta N_{c}}{4\pi ^{2}}~-\frac{\lambda
^{2}N_{f}}{4\pi ^{2}}=2\frac{\zeta }{\lambda }\beta _{\lambda },
\]
so that $\beta _{u}=0$.

The value (\ref{zita}) is large, because of the $\Delta $ in the
denominator. This means that the perturbative expansion in powers of the
energy is meaningful if the energy is much smaller than the effective Planck
scale
\begin{equation}
M_{P\mathrm{eff}}=\frac{1}{\kappa _{\mathrm{eff}}}=\frac{\Delta }{\kappa }.
\label{defl}
\end{equation}
The other factor in (\ref{zita}) can be taken of order one, if $\lambda
^{2}(\mu )N_{c}$ is kept fixed in the large $N_{c},N_{f}$ limit.

\bigskip

Repeating the same calculation for the four-fermion counterterms, the result
is
\begin{eqnarray*}
\frac{\xi _{1}N_{c}}{16\pi ^{2}} &=&\frac{16\Delta }{225N_{c}}\lambda
^{2},\qquad \frac{\xi _{2}N_{c}}{16\pi ^{2}}=-\frac{56\Delta }{225N_{c}}%
\lambda ^{2},\qquad \frac{\xi _{3}N_{c}}{16\pi ^{2}}=-\frac{92\Delta }{225}%
\lambda ^{2},\qquad \frac{\xi _{4}N_{c}}{16\pi ^{2}}=\frac{52\Delta }{225}%
\lambda ^{2}, \\
\frac{\lambda _{1}N_{c}}{16\pi ^{2}} &=&\frac{544\Delta }{3225N_{c}}\lambda
^{2},\qquad \frac{\lambda _{2}N_{c}}{16\pi ^{2}}=\frac{8\Delta }{43N_{c}}%
\lambda ^{2},\qquad \frac{\lambda _{3}N_{c}}{16\pi ^{2}}=-\frac{94\Delta }{%
3225}\lambda ^{2},\qquad \frac{\lambda _{4}N_{c}}{16\pi ^{2}}=\frac{2\Delta
}{43}\lambda ^{2}, \\
\frac{\eta _{1}N_{c}}{16\pi ^{2}} &=&-\frac{N_{c}\Delta }{45}\lambda
^{2},\qquad \frac{\eta _{2}N_{c}}{16\pi ^{2}}=\frac{4\Delta }{675}\lambda
^{2}.
\end{eqnarray*}
Here no denominator contains $\Delta $, so (\ref{defl}) is not modified.

\section{Applications to supersymmetric theories}

In this section I prove that the non-renormalization theorem for the chiral
operators in supersymmetric theories does not preclude the existence of
solutions of the quasi-finiteness equations. I construct finite and
quasi-finite chiral irrelevant deformations.

Consider a supersymmetric theory in four dimensions, formulated using N=1
superfields. A well-known non-renormalization theorem states that no chiral
counterterm appears in the renormalization of the theory. A chiral
counterterm is a term that cannot be written as the integral in $\mathrm{d}%
^{4}\theta $ of a local superfield. Consider for example the chiral
lagrangian term
\begin{equation}
\mathit{C}_{k}=Y_{k}\int \Phi ^{k}~\mathrm{d}^{2}\theta ,  \label{chiral}
\end{equation}
where $Y_{k}$ is a coupling constant and $\Phi $ is an elementary chiral
superfield. This operator has level $k-3$. The non-renormalization theorem
implies that
\begin{equation}
Y_{k}\int \Phi ^{k}~\mathrm{d}^{2}\theta =Y_{k}Z_{k}Z_{\Phi }^{k/2}\int \Phi
^{k}~\mathrm{d}^{2}\theta ,\qquad \text{or }Z_{k}Z_{\Phi }^{k/2}=1,\text{
i.e. }\gamma _{(k)}=k\gamma _{\Phi }\text{.}  \label{gh}
\end{equation}
Thus, the anomalous dimension $\gamma _{(k)}$ of a chiral operator of the
form (\ref{chiral}) is $k$ times the anomalous dimension of the elementary
field $\Phi $. Moreover, the equation $Y_{k\mathrm{B}}=Y_{k}Z_{k}=Y_{k}Z_{%
\Phi }^{-k/2}$ implies that the beta function is
\[
\beta _{(k)}=kY_{k}\gamma _{\Phi }=Y_{k}\gamma _{(k)}.
\]
This means that the quantity $\delta $ of formula (\ref{betagel}) vanishes,
i.e. that the chiral operators are protected, in the sense explained in
section 2.

\bigskip

Now, consider an interacting superconformal field theory. Examples are the
IR fixed points of certain UV-free theories (see \cite{noi} for a collection
of properties of these conformal theories). Turn on the deformation $\mathit{%
C}_{k}$, which has level $\ell =k-3$. The requirement (\ref{dimma}) for the
existence of solutions of the quasi-finiteness equations reads
\[
\gamma _{(nk-3n+3)}\neq n\gamma _{(k)}
\]
for every integer $n>1$. Using (\ref{gh}) this condition becomes
\[
(nk-3n+3)\gamma _{\Phi }\neq kn\gamma _{\Phi },
\]
which is always true if $\gamma _{\Phi }\neq 0$. The value of $\gamma _{\Phi
}$ is known in various models, using the NSVZ exact beta function \cite{nsvz}%
. Typically, $\gamma _{\Phi }\neq 0$ in N=1 superconformal field theories,
but there exist finite N=2 and N=4 supersymmetric theories where $\gamma
_{\Phi }=0$. Since, however, the chiral operators are protected, the fact
that (\ref{dimma}) is violated is not a problem for the construction of
consistent irrelevant deformations. I consider the cases $\gamma _{\Phi
}\neq 0$ and $\gamma _{\Phi }=0$ separately.

\bigskip

I recall from section 2 that an irrelevant deformation (\ref{finitesol}) is
made of a lowest-level operator $\mathcal{O}_{\ell }$ and a queue. The
lowest-level term is multiplied by an arbitrary parameter $\widetilde{\kappa
}$ that can run according to (\ref{run}).\ The queue consists of infinitely
many lagrangian terms, whose couplings are determined by the
quasi-finiteness equations. I\ call \textit{chiral} a deformation whose
lowest-level term $\mathcal{O}_{\ell }$ is chiral. Now, repeat the
construction of section 2 for a chiral deformation. If $\gamma _{\Phi }\neq
0 $, the parameter $\widetilde{\kappa }$ still runs. Moreover, the queue
does not contain other chiral terms, because their coefficients are set to
zero by the quasi-finiteness equations.

Now, consider a case where $\gamma _{\Phi }=0$, for example N=4
supersymmetric Yang-Mills theory. In the formalism of N=1 superfields, the
theory contains a vector multiplet and three chiral multiplets $\Phi ^{i}$.
The fields $\Phi ^{i}$ have zero anomalous dimensions and therefore the
chiral operators, for example
\[
\int Y_{i_{1}\cdots i_{n}}\Phi ^{i_{1}}\cdots \Phi ^{i_{n}}~\mathrm{d}%
^{2}\theta ,
\]
are finite and protected. (Here $Y_{i_{1}\cdots i_{n}}$ is a constant
tensor.) Consider a chiral irrelevant deformation of N=4 supersymmetric
Yang-Mills theory. Since the anomalous dimension of the lowest-level
operator is zero by assumption, the running (\ref{run}) is trivial, i.e. $%
\widetilde{\kappa }$ is an arbitrary finite parameter. The deformation is
therefore finite. Since chiral operators are protected, the coefficients of
the other chiral terms in the queue can be consistently set to zero. If the
queue of a chiral deformation contains no other chiral term, the chiral
deformation is \textit{simple}. If other chiral terms appear in the queue
(multiplied by arbitrary parameters), then it is a \textit{multiple} chiral
deformation. The queue contains also non-chiral terms. For these, the
condition (\ref{dimma}) is non-trivial. Since the lowest-level term is
finite ($r_{\ell }=0$), the condition (\ref{dimma}) for the existence of the
deformation states that the non-chiral operators of levels $n\ell $ should
have non-vanishing anomalous dimensions. This is generically true. Then, the
quasi-finiteness equations (which are actually \textit{finiteness}
equations, in this particular case) admit a solution. The perturbative
expansion in powers of the energy is well-defined if the quantity $\eta $
defined in eq. (\ref{bound1}) is strictly positive. Since the anomalous
dimensions of non-chiral operators are generically non-trivial already at
the first loop order, $\eta $ is typically of order $g^{2}$, where $g$ is
some gauge coupling, and the ``effective Planck mass'' $M_{P\mathrm{eff}%
}=1/\kappa _{\mathrm{eff}}$ is typically of order $\sim
g^{2}M_{P}=g^{2}/\kappa $.

\bigskip

In conclusion, I have proved that the requirement (\ref{dimma})
for the existence of quasi-finite irrelevant deformations is not
in contradiction with the non-renormalization theorem of
supersymmetric theories. The construction of quasi-finite
irrelevant deformations of N=1 superconformal field theories
proceeds as in the absence of supersymmetry, while N=2 and N=4
finite theories admit also finite chiral irrelevant deformations.

Observe that the finite chiral irrelevant deformations of superconformal
field theories are also good examples of finite four-dimensional
power-counting non-renormalizable theories. They can be seen as particular
applications of the construction elaborated in ref. \cite{mid}. Their
renormalization requires only field redefinitions, but no coupling
redefinition.

\section{Conclusions}

Using the strategy of this paper, it is possible to construct consistent
irrelevant deformations of interacting conformal field theories. These
deformations have a scale, which multiplies the lowest-level irrelevant
term, a finite number of coupling constants, and a queue made of infinitely
many lagrangian terms. A finite number of renormalization constants, plus
field redefinitions, are sufficient to remove the divergences. The scale can
run (quasi-finite deformations) or not (finite deformations). If the scale
runs, the queue of the deformation runs coherently with the lowest-level
term. In certain families of supersymmetric theories it is possible to
construct chiral finite deformations.

The perturbative expansion is meaningful for energies much smaller than an
effective Planck mass. The effective Planck mass becomes small when the
interaction of the renormalizable sector of the theory becomes weak.

Generalizations are possible. For example, in some cases it is possible to
construct irrelevant deformations of running power-counting renormalizable
theories. However, this issue is technically more tricky and is left for a
future publication.

Although the ideas of this paper do not apply directly to quantum gravity, a
more general framework where they do might exist. Then, the quantization of
gravity might be possible only thanks to the existence of other interacting
matter. The effects of quantum gravity might show up at energies some orders
of magnitude smaller than the Planck mass, depending on the strength of the
interaction in the matter sector.

The results of this paper and ref. \cite{mid} suggest that power-counting
non-renormalizable theories are candidate to play a relevant role in the
description of fundamental physics. Certainly, the problem of predictivity
of fundamental field theory needs to be carefully reconsidered in the light
of these results.

\section{Appendix. Useful identities}

\setcounter{equation}{0}

In this appendix I collect useful identities, fields equations and certain
manipulations that are helpful to identify the irreducible set of irrelevant
terms, and to simplify the counterterms for the study of the Pauli
deformation of the IR fixed point of massless non-Abelian Yang-Mills theory
with $N_{c}$ colors and $N_{f}\lesssim 11N_{c}/2$ flavors.

\bigskip

\textbf{Notation and identities for gauge group and representations. }The
notation is such that $[T^{a},T^{b}]=f^{abc}T^{c}$, $(f^{abc})^{*}=f^{abc}$,
$(T^{a})^{\dagger }=-T^{a}$. Useful identities are
\begin{equation}
\mathrm{tr}[T^{a}T^{b}]=-\frac{1}{2}\delta ^{ab},\qquad
T_{ij}^{a}T_{kl}^{a}=-\frac{1}{2}\delta _{il}\delta _{jk}+\frac{1}{2N_{c}}%
\delta _{ij}\delta _{kl}.  \label{id1}
\end{equation}

\bigskip

\textbf{Useful identities for the Dirac algebra.} The products between the
elements 1, $\gamma _{5}=\varepsilon ^{\mu \nu \rho \sigma }\gamma _{\mu
}\gamma _{\nu }\gamma _{\rho }\gamma _{\sigma }/4!$, $\gamma _{\mu }$, $%
\gamma _{5}\gamma _{\mu }$, $\sigma _{\mu \nu }=-i[\gamma _{\mu },\gamma
_{\nu }]/2$ of the Clifford algebra are immediate or can be read from the
following table:
\begin{eqnarray*}
\sigma _{\mu \nu }\gamma _{\rho } &=&i\delta _{\mu \rho }\gamma _{\nu
}-i\delta _{\nu \rho }\gamma _{\mu }+i\varepsilon _{\mu \nu \rho \sigma
}\gamma _{\sigma }\gamma _{5},\qquad \gamma _{\rho }\sigma _{\mu \nu
}=-i\delta _{\mu \rho }\gamma _{\nu }+i\delta _{\nu \rho }\gamma _{\mu
}+i\varepsilon _{\mu \nu \rho \sigma }\gamma _{\sigma }\gamma _{5}, \\
\sigma _{\mu \nu }\gamma _{5} &=&-\frac{1}{2}\varepsilon _{\mu \nu \alpha
\beta }\sigma _{\alpha \beta },\qquad \qquad \qquad \qquad \qquad \gamma
_{\mu }\gamma _{\nu }=\delta _{\mu \nu }+i\sigma _{\mu \nu }, \\
\sigma _{\mu \nu }\sigma _{\alpha \beta } &=&\delta _{\mu \alpha }\delta
_{\nu \beta }-\delta _{\mu \beta }\delta _{\nu \alpha }-\varepsilon _{\mu
\nu \alpha \beta }\gamma _{5}-i(\delta _{\nu \alpha }\sigma _{\mu \beta
}-\delta _{\nu \beta }\sigma _{\mu \alpha }+\delta _{\mu \beta }\sigma _{\nu
\alpha }-\delta _{\mu \alpha }\sigma _{\nu \beta }).
\end{eqnarray*}

\bigskip

\textbf{Field equations. }The $\mathcal{O}(\kappa )$-field equations read
\begin{eqnarray}
D\!\!\!\!\slash_{ij}\Psi _{j}^{I}+\kappa \lambda Z_{\lambda
}T_{ij}^{a}\sigma _{\mu \nu }\Psi _{j}^{I}\mathcal{F}_{\mu \nu }^{a}+%
\mathcal{O}(\kappa ^{2}) &=&0,  \label{field1} \\
\frac{\mu ^{-\varepsilon }}{g^{2}Z_{g}^{2}}\mathcal{D}_{\nu }^{ab}\mathcal{F}%
_{\mu \nu }^{b}+\overline{\Psi }_{i}^{I}\gamma _{\mu }T_{ij}^{a}\Psi
_{j}^{I}+2\kappa \lambda Z_{\lambda }\mathcal{D}_{\nu }^{ab}(\overline{\Psi }%
_{i}^{I}\sigma _{\mu \nu }T_{ij}^{a}\Psi _{j}^{I})+\mathcal{O}(\kappa ^{2})
&=&0.  \label{field2}
\end{eqnarray}

\bigskip

\textbf{Identities for terms of level }$\mathbf{1}$\textbf{.} Here I write
some identities that are useful for the simplification of the $\mathcal{O}%
(\kappa )$ counterterms. The first one is obvious
\begin{equation}
\int \overline{\psi }D^{2}\psi =\int \overline{\psi }D\!\!\!\!\slash^{2}\psi
-\frac{i}{2}\int \overline{\psi }T^{a}\sigma _{\mu \nu }\psi ~F_{\mu \nu
}^{a}.  \label{one}
\end{equation}
The second one is obtained dropping the $\Box $-terms on each side:
\begin{equation}
\int \overline{\psi }T^{a}\psi ~\partial _{\mu }A_{\mu }^{a}+2\int (\partial
_{\mu }\overline{\psi })T^{a}\psi ~A_{\mu }^{a}=-\int A_{\mu }^{a}~\overline{%
\psi }\left( \gamma _{\mu }T^{a}D\!\!\!\!\slash-\overleftarrow{D\!\!\!\!%
\slash}T^{a}\gamma _{\mu }\right) \psi +\frac{i}{2}\int \overline{\psi }%
T^{a}\sigma _{\mu \nu }\psi ~F_{\mu \nu }^{a},  \label{olb}
\end{equation}
up to terms $A$-$A$-$\overline{\psi }$-$\psi $. The integral appears to
allow partial integrations and drop total derivatives. It is understood that
the derivative covered with a left arrow acts only on $\overline{\psi }$.

The identities (\ref{one}) and (\ref{olb}) are useful to express certain
counterterms as sums of objects proportional to the $\mathcal{O}(\kappa
^{0}) $-field equations plus the Pauli term.

\bigskip

\textbf{Identities for terms of level }$\mathbf{2}$\textbf{.} A similar work
has to be done at $\mathcal{O}(\kappa ^{2})$. It is sufficient to work out
identities up to terms of the form $A$-$A$-$\overline{\psi }$-$\psi $, which
are unnecessary for the computations of the paper. Moreover, in the
manipulations of the $\mathcal{O}(\kappa ^{2})$ counterterms, the terms
proportional to the $\mathcal{O}(\kappa ^{0})$-field equations can be safely
dropped. This operation is denoted with an arrow.

With these conventions, it is easy to show that
\[
0\leftarrow \int F_{\mu \nu }^{a}~\overline{\psi }\left( T^{a}\sigma _{\mu
\nu }D\!\!\!\!\slash+\overleftarrow{D\!\!\!\!\slash}\sigma _{\mu \nu
}T^{a}\right) \psi =2i\int F_{\mu \nu }^{a}~\overline{\psi }T^{a}\gamma
_{\nu }\left( \partial _{\mu }-\overleftarrow{\partial _{\mu }}\right) \psi
.
\]
To derive this identity it is sufficient to partially integrate, use the
Bianchi identity and ignore the terms of the form $A$-$A$-$\overline{\psi }$-%
$\psi $.

Furthermore, partially integrating and using the gauge-field equations, we
have also
\[
2i\int F_{\mu \nu }^{a}~\overline{\psi }T^{a}\gamma _{\nu }\left( \partial
_{\mu }+\overleftarrow{\partial _{\mu }}\right) \psi \rightarrow
-2ig^{2}\int \left( \overline{\psi }T^{a}\gamma _{\mu }\psi \right) ^{2}.
\]
Combining the two, we get
\begin{equation}
\int F_{\mu \nu }^{a}~(D_{\mu }\overline{\psi })T^{a}\gamma _{\nu }\psi
\rightarrow -\frac{1}{2}g^{2}\int \left( \overline{\psi }T^{a}\gamma _{\mu
}\psi \right) ^{2},  \label{uno}
\end{equation}
up to $\mathcal{O}(\kappa )$, terms $A$-$A$-$\overline{\psi }$-$\psi $ and
terms proportional to the $\mathcal{O}(\kappa ^{0})$-field equations.

Similarly, we have
\begin{eqnarray*}
0\leftarrow \int F_{\mu \nu }^{a}~\overline{\psi }\left( T^{a}\sigma _{\mu
\nu }D\!\!\!\!\slash-\overleftarrow{D\!\!\!\!\slash}\sigma _{\mu \nu
}T^{a}\right) \psi =i\int \varepsilon _{\mu \nu \rho \sigma }F_{\mu \nu
}^{a}~\overline{\psi }T^{a}\gamma _{\sigma }\gamma _{5}\left( \partial
_{\rho }-\overleftarrow{\partial _{\rho }}\right) \psi &+& \\
-2i\int D_{\mu }^{ab}F_{\mu \nu }^{b}~\overline{\psi }T^{a}\gamma _{\nu
}\psi \rightarrow i\int \varepsilon _{\mu \nu \rho \sigma }F_{\mu \nu }^{a}~%
\overline{\psi }T^{a}\gamma _{\sigma }\gamma _{5}\left( \partial _{\rho }-%
\overleftarrow{\partial _{\rho }}\right) \psi -2ig^{2}\int \left( \overline{%
\psi }T^{a}\gamma _{\mu }\psi \right) ^{2}. &&
\end{eqnarray*}
Combining this formula with a simple consequence of the Bianchi identity,
namely
\[
0=i\int \varepsilon _{\mu \nu \rho \sigma }F_{\mu \nu }^{a}~\overline{\psi }%
T^{a}\gamma _{\sigma }\gamma _{5}\left( \partial _{\rho }+\overleftarrow{%
\partial _{\rho }}\right) \psi ,
\]
we obtain
\begin{equation}
\int \varepsilon _{\mu \nu \rho \sigma }F_{\mu \nu }^{a}~(D_{\rho }\overline{%
\psi })T^{a}\gamma _{\sigma }\gamma _{5}\psi \rightarrow -g^{2}\int \left(
\overline{\psi }T^{a}\gamma _{\mu }\psi \right) ^{2},  \label{due}
\end{equation}
up to $\mathcal{O}(\kappa )$, terms $A$-$A$-$\overline{\psi }$-$\psi $ and
terms proportional to the $\mathcal{O}(\kappa ^{0})$-field equations.

\end{document}